\documentclass[letterpaper,twocolumn,10pt]{article}
\PassOptionsToPackage{hyphens}{url}
\usepackage{usenix-2020-09}

\def\isanonymous{0}
\def\issmallscreen{0}

\usepackage[l2tabu,orthodox]{nag}

\usepackage{ifxetex}

\usepackage{microtype}
\usepackage{booktabs,caption}
\usepackage{hyphenat}
\usepackage[flushleft]{threeparttable}
\usepackage{ifthen}
\newcommand{\anonymous}[2]{%
	\ifthenelse{\equal{\isanonymous}{1}}%
	{{#1}}%
	{{#2}}%
}

\newcommand{\smallscreen}[2]{%
	\ifthenelse{\equal{\issmallscreen}{1}}%
	{{#1}}%
	{{#2}}%
}

\ifthenelse{\equal{\issmallscreen}{1}}{
	\usepackage{geometry}
	\geometry{margin=5cm}
	\setlength{\marginparwidth}{2cm}
}{}

\usepackage{xcolor}
\definecolor{oxygenorange}{HTML}{FFDD00}
\usepackage[color=oxygenorange]{todonotes}

\ifthenelse{\equal{\isanonymous}{1}}
{
	\newcommand{\rikke}[2][]{}
	\newcommand{\jess}[2][]{}
}{
	\newcommand{\rikke}[2][inline]{\todo[#1]{\textbf{rikke:} #2}\xspace}
	\newcommand{\jess}[2][inline]{\todo[#1]{\textbf{jess:} #2}\xspace}
}

\usepackage{amsmath,amsfonts}  
\usepackage{xspace}

\usepackage[lambda,landau,operators,probability,sets,logic,complexity,asymptotics]{cryptocode}

\usepackage{booktabs}  
\usepackage{comment}
\usepackage{enumitem}

\usepackage{navigator}
\usepackage{url}
\usepackage{graphicx}
\usepackage[export]{adjustbox}
\usepackage{float}
\usepackage{multirow}
\usepackage{longtable}
\usepackage{array}
\newcolumntype{L}{>{\raggedright\arraybackslash}p{3cm}}
\usepackage{pdfpages}
\usepackage{subcaption}   
\usepackage{tikz,pgfplots,pgfplotstable}
\usepgfplotslibrary{statistics}
\usetikzlibrary{calc}
\usetikzlibrary{arrows}
\usetikzlibrary{positioning}

\pgfplotsset{
	tick label style={font=\small},
	label style={font=\small},
	legend style={font=\small, cells={anchor=west}}
}

\definecolor{DarkPurple}{HTML}{332288}
\definecolor{DarkBlue}{HTML}{6699CC}
\definecolor{LightBlue}{HTML}{88CCEE}
\definecolor{DarkGreen}{HTML}{117733}
\definecolor{DarkRed}{HTML}{661100}
\definecolor{LightRed}{HTML}{CC6677}
\definecolor{LightPink}{HTML}{AA4466}
\definecolor{DarkPink}{HTML}{882255}
\definecolor{LightPurple}{HTML}{AA4499}

\definecolor{DarkBrown}{HTML}{604c38}
\definecolor{DarkTeal}{HTML}{23373b}
\definecolor{LightBrown}{HTML}{EB811B}
\definecolor{LightGreen}{HTML}{14B03D}

\usepackage{listings}
\lstdefinelanguage{Sage}[]{Python}{morekeywords={True,False,sage,cdef,cpdef,ctypedef,self},sensitive=true}
\begin{document}
	
	\title{Othered, Silenced and Scapegoated: \\Understanding the Situated Security of Marginalised Populations in Lebanon}
	\anonymous{
		\author{}
	}{
		\author{
			{\rm Jessica McClearn}\\
			Royal Holloway, University of London\\
			jessica.mcclearn.2021@live.rhul.ac.uk
			\and
			{\rm Rikke Bjerg Jensen}\\
			Royal Holloway, University of London\\
			rikke.jensen@rhul.ac.uk
			\and
			{\rm Reem Talhouk}\\
			Northumbria University\\
			reem.talhouk@northumbria.ac.uk
		}
		
	}

	\clubpenalty=1
	\displaywidowpenalty=1
	\widowpenalty=1
	
	\maketitle

\begin{abstract}
In this paper we explore the digital security experiences of marginalised populations in Lebanon such as LGBTQI+ identifying people, refugees and women. We situate our work in the post-conflict Lebanese context, which is shaped by sectarian divides, failing governance and economic collapse. We do so through an ethnographically informed study conducted in Beirut, Lebanon, in July 2022 and through interviews with 13 people with Lebanese digital and human rights expertise. Our research highlights how LGBTQI+ identifying people and refugees are \emph{scapegoated} for the failings of the Lebanese government, while women who speak out against such failings are \emph{silenced}. We show how government-supported incitements of violence aimed at transferring blame from the political leadership to these groups lead to amplified digital security risks for already at-risk populations. Positioning our work in broader sociological understandings of security, we discuss how the Lebanese context impacts identity and ontological security. We conclude by proposing to design for and with positive security in post-conflict settings. 
\end{abstract}

\section{Introduction}\label{sec:introduction}
Experiences of political unrest, corruption and economic crisis shape the daily lives of people living in post-conflict Lebanon~\cite{AMAA2022,BouElA2016,abi2018,verdeil2018}. Post-conflict societies are characterised by being in a `transition continuum', rather than in a fixed state of conflict or peace. Here, key transition milestones include ending violence, a peace agreement, disarmament, functioning governance, societal reconciliation and economic recovery~\cite{brown2011typology}. The Lebanese Civil War (1975-1990) ended with the Al Taif Agreement, but the legacies of conflict have moved from (mostly) armed conflict to manifesting in Lebanese political and governance structures. A history of conflict rooted in fundamental disagreements between Christian and Muslim Lebanese on the historicity of the country and their visions for Lebanese identity has led to society-wide fragmentation~\cite{Salibi1990}. Identity and security in Lebanon are linked to, and through, such sectarian divides, where distinctions between \emph{them} and \emph{us} dominate public discourse.\footnote{In `sectarianism', belonging to a religious sect is the main mechanism for legal recognition, political participation and, thus, social control~\cite{Mikdashi2022}.} 

The corrupt Lebanese regime whose politics functions along sectarian divides has flung the country into one of the world's worst economic crises in 150 years~\cite{worldbank2021}. In 2021, the OHCHR reported that Lebanese political and financial leaders were ``responsible for forcing most of the country's population into poverty''~\cite{OHCHR2022}, while the World Bank labelled the economic collapse in the country a ``ponzi scheme''~\cite{worldbank2022} leading to a ``deliberate depression''~\cite{worldbank2020}. Stories of how Lebanese people have been locked out of their bank deposits have been covered in international media~\cite{press:lebanesebanks:reuters}, with some reporting that ``three-quarters of the Lebanese population are in poverty''~\cite{press:Lebanon:FT} and that 80\% of the Lebanese population have no access to basic rights such as education and healthcare~\cite{HRW2021b}. Post-conflict legacies thus underpin our work on digital security in Lebanon, where sectarian tensions also manifest online.  

Our work shows how marginalised groups in Lebanon, particularly LGBTQI+ identifying people and refugees, are \emph{scapegoated} for Lebanese government failures, while women who speak out against such failings or challenge patriarchal norms experience amplified \emph{silencing}. We refer to these practices as \emph{othering}. We ground our findings in related digital security scholarship that has identified a diverse range of risks experienced by, e.g., refugees and migrants~\cite{CHI:JenColTal20,SP:SLIRK18} and LGBTQI+ identifying people~\cite{USENIX:GHRR22,CHI:LHKZH20,CHI:GMSMTS18} in higher-risk contexts. Further, existing work has highlighted how the ability to secure digital access is of particular concern for displaced populations and is often intermittent~\cite{tachtler2021unaccompanied,sabie2019moving}. Such work shines a light on how some populations become higher-risk due to their socio-economic status, political convictions or gender identity. Further, we position our work within wider conceptualisations of security and identity; particularly sociological concepts of ontological security, e.g.~\cite{giddens1991,mcsweeney1999security}, and positive security, e.g.~\cite{roe2008value,mcsweeney1999security}, to highlight the need for digital security research to consider the intersecting forms of marginalisation and societal structures of inequity and insecurity. 

\paragraph{Contributions.} This work draws on ethnographically informed fieldwork over two weeks in Beirut, Lebanon, in July 2022. As well as observation \emph{in situ}, the research involved 12 interviews with 13 people who had expertise in digital and human rights in Lebanon. The aim of the research was to explore how digital security is shaped by the Lebanese post-conflict context and the societal-wide failing infrastructures. We report on findings pertaining to notions of othering, scapegoating and silencing targeted at marginalised people. By positioning our work within digital security, we place the security of technology in the context of people and society. 

Our work makes four distinct contributes to digital security scholarship. First, we show how already marginalised populations in Lebanon are targeted through practices of othering that are actively supported by the Lebanese sectarian leadership. This othering appears on a scale ranging from online silencing, to targeted abuse and hatred, to outright scapegoating. Our findings highlight how these practices amplify the digital security risks of those being targeted, particularly women, LGBTQI+ identifying people and refugees. We further show how the Lebanese government uses othering to shift blame for the political and economic crisis onto these marginalised populations. Our findings demonstrate how these forms of othering are rooted in perceptions that marginalised groups morally challenge the traditional politics that maintains the sectarian Lebanese regime, its state power and societal position. Second, we show how the failures of institutions within a post-conflict setting contribute to the fragmentation of Lebanese identities and allow for the normalisation of an `us' and `them' discourse, which further reinforces practices of othering and how they manifest online. Third, we draw on positive security for design interventions that enhance dialogue and reconciliation within post-conflict contexts. We show that to succeed in this endeavour, the security experiences of marginalised populations need to be identified through situated research that understands the enmeshed -- social and technological -- nature of digital security. Fourth, we conducted an ethnographically informed study that brings to the fore the situated security of the populations under study, as well as our interpretations of it. In doing so, we also contribute to a diversification of the methodological approaches employed to understand digital security. 

\section{Related Work}\label{sec:related-work}
Our work speaks to broader conceptualisations of security, which we set out in this section. First, in Section~\ref{sec:rw-ontological-security}, we bring ontological security into conversation with identity work, before engaging with the concept of scapegoating as it pertains to notions of blame and othering. We do so in Section~\ref{sec:rw-scapegoating}. In Section~\ref{sec:rw-higher-risk} we engage with existing research on higher-risk populations. Collectively, these works lay the conceptual foundations for our presentation of findings in Section~\ref{sec:findings} and discussion in Section~\ref{sec:discussion}.

\subsection{Ontological Security and Identity}\label{sec:rw-ontological-security}
We ground and deepen our discussion on the practices of othering that we observed in our findings by situating it within \textbf{ontological security}\footnote{The concept of \emph{ontological security} was developed by sociologist Anthony Giddens~\cite{giddens1991}.} as it relates to identity and positive security. Practices of othering create a distinction between `us' and `them' and is a process inherently linked to ontological security and the securing of the self. Concretely, we use sociologist Bill McSweeney's~\cite{mcsweeney1999security} framing of security as encompassing both the freedom to live without fear and protection from harm. We thus understand ontological security as the creation of a sense of security through the leveraging of trusted relations and routines~\cite{croft2012constructing,croft2017fit}. Ontological security is the feeling that one's understanding of and position within the world is stable and reliable, providing a sense of predictability~\cite{roe2008value,mcsweeney1999security}. This definition also ties in with Steele's~\cite{steele2005} understanding of ontological security as ``security as being'' and Mitzen's~\cite{mitzen2006} ``security of the self'', foregrounding the relationship between security and identity. We use ontological security as a framework for interrogating how security and identity are connected in the Lebanese context.

 Identity is not fixed but is continuously shaping -- and shaped by -- how people relate to each other and to the world that surrounds them. Identity therefore is a dialogical practice through which people establish a sense of ontological security ~\cite{cunliffe2001}. While the relationship between ontological security and identity work has many dimensions, essentially ontological security may lead to particular forms of identity work. For example, constructing and/or negotiating an identity that holds some form of power or ensures a sense of belonging may lead to greater ontological security. Recent HCI scholarship has also drawn on identity work as understood within ontological security. For example, the authors of~\cite{schoemaker2021identity} showed how refugees in Jordan, Lebanon and Uganda negotiated their digital identities within the UNHCR identity management system to maximise access to services and, in turn, their ontological security. In~\cite{CHI:DosSem18,CHI:DosSem19}, Dosono and Semaan referred to ``identity work as deliberation'' when exploring identity work among Asian American and Pacific Islander communities on Reddit during the 2016 US Presidential Election. The concept of ontological security has also been employed in research on isolated communities at the margins of society~\cite{CHI:JCWL20} and refugees (re)settling in and accessing a new country~\cite{CHI:ColJen19,CHI:ColJenTal18}. These works raise the question of what kind of security is needed to better cater to marginalised populations whose threat horizons are amplified through digital technology~\cite{CHI:ColJenTal18}. 

Similar to ontological security, \textbf{positive security} builds on McSweeney's~\cite{mcsweeney1999security} work where positive security is the ability to pursue one's interests and fulfil one's needs~\cite{gjorv2012security} through trusted relations. Indeed, McSweeney was one of the first to acknowledge the significance of positive security, arguing that the noun \emph{security} refers to an object to be protected, but that the adjective \emph{secure} refers to making something possible~\cite[p.14]{mcsweeney1999security}. This is distinct from \textbf{negative security}, where security is gained through the protection from threats~\cite{gjorv2012security}. McSweeney~\cite{mcsweeney1999security} further articulated how people need day-to-day routines to establish a sense of self and be able to relate to others as central to positive security. Building on McSweeney's~\cite{mcsweeney1999security} work, Roe~\cite[p.778]{roe2008value} notes that ``[p]ositive security thus relates to the securities and insecurities that individuals, and the communities in which people live, routinely create for one another.'' A positive security lens thus enables us to identify the barriers and threats faced by individuals and groups when attempting to secure their aspirations and ambitions, i.e. making something possible. In Section~\ref{sec:dis-post-conflict} we consider how designing for positive security in post-conflict settings requires design interventions that focus on dialogue and building trust relations across fragmented societies.

\subsection{Scapegoating: Blame and Othering}\label{sec:rw-scapegoating}
The use of scapegoating can be seen as a mechanism of restoring order and reinforcing group identity~\cite{joffe1999risk}. Scapegoating, or the threat of it, can thus be seen as a way to control someone's identity through processes of othering. International relations and statecraft literature has long considered the power of strategic scapegoating to control populations through the use of diversionary theory~\cite{gent2009scapegoating,foster2010rallies} or through the link of violence and scapegoat ideology~\cite{thomas2015rethinking}. Despite this work providing insight into how populations have been painted as scapegoats little research exists into the relationship between security, marginalised populations and forms of othering.

Through the practice of scapegoating `the other' is identified as `not me', which Mythen~\cite[p.101]{mythen2004ulrich} articulates as ``the not me -- other'' approach and highlighting it as a convenient way to dispatch blame towards targeted groups. Othering and scapegoating are thus often connected and surface in times of crisis, where they function as methods used to assign blame~\cite{douglas2002scapegoats}. This is also reinforced by Joffe~\cite{joffe1999risk}, who argues that the continuous degradation of `others' across societies is a mechanism of assigning blame and ``becomes magnified at times of crisis''. Security scholar Neocleous~\cite{neocleous2008} explores the role of othering in relation to identity construction and fear of threats, which are often grounded in feelings of uncertainty. Here, the production of `others' is often amplified through government and media discourses, which can reproduce negative stereotypes about `the other' and ``apportionment of blame, masking the multicausal reproduction of risk''~\cite{mythen2004ulrich}. For the purpose of this study we understand scapegoating as ``the act of blaming an out-group when the in-group experiences frustration or is blocked from obtaining a goal''~\cite{allport1954nature}. 

\subsection{Security for Higher-Risk Populations}\label{sec:rw-higher-risk}
Section~\ref{sec:rw-ontological-security} focused on ontological and positive security, while Section~\ref{sec:rw-scapegoating} foregrounded practices of othering. Here, we further contextualise our contributions by drawing on the growing body of security-driven work that has focused on higher-risk and marginalised populations, recognising that their security needs are not well served by the technologies they rely upon~\cite{USENIX:ABJM21,USENIX:GHRR22}. The participants in our study can be considered \emph{higher-risk} in multiple ways, where their situated security experiences of living through conflict also heighten their digital security needs. This intertwining of technological and societal-level security is exemplified in existing work with higher-risk populations. Research with migrants and refugees has identified how language and local customs or social structures, for example, become barriers to them fulfilling their security needs in different contexts and in addition to technological obstacles~\cite{CHI:ColJenTal18,CHI:JenColTal20,SP:SLIRK18,CHI:GMSMTS18}. These bodies of work highlight how unmet digital security needs lead to a series of informal and socially rooted practices that aim to mitigate experiences of ontological insecurity. Other groups of higher-risk users, such as political activists, have been shown to rely on the digital security practices performed by their social relations, such as other activists, as well as the security of the technology they use~\cite{SP:DSKB21,ErmHalMus17,CHI:BSCU21,USENIX:ABJM21}. This work highlights the significance of trusted relations for ontological security that translate to digital security practices. Recent work on LGBTQI+ identifying people has also shown how they look for support from trusted queer groups to navigate questions of identity, personal safety and security~\cite{USENIX:GHRR22,CHI:LHKZH20}. Without doing so explicitly, these works speak to the enmeshed nature of ontological security and identity, as we note in Section~\ref{sec:rw-ontological-security}; thus suggesting how the ontological security needs of marginalised populations manifests through digital security practices.

The intertwining of technological and societal security is further highlighted in research on higher-risk populations. For example, the authors of~\cite{sambasivan2018privacy} show how South Asian women developed performative practices to protect their privacy. In~\cite{matthews2017stories}, the authors demonstrate how survivors of intimate partner abuse develop distinct privacy practices that pertain to the different stages of their experiences of abuse. This work speaks to a broader body of scholarship that has focused on the security and privacy needs of those living with intimate partner violence, e.g.~\cite{freed2017digital,freed2019my,tseng2021digital}. Further, research on the protection mechanisms developed by migrant domestic workers highlights how these are grounded in online and offline support networks~\cite{USENIX:SCBAPB22}. Finally, tying together notions of financial insecurity and digital security, the authors of~\cite{sleeper2019tough} explored how pressures related to homelessness and limited financial resources impacted people's security and privacy practices. The literature referred to here shows how marginalisation shapes digital security in different ways, while being rooted in wider societal contexts.

\section{Methodology}\label{sec:methodology}
Here, we outline the research design, including participant recruitment and the interview process as well as the ethical considerations that guided our research. We do so in Section~\ref{sec:research-design}. In Section~\ref{sec:data-analysis}, we discuss our data analysis and researcher positionalities. 

\subsection{Research Design}\label{sec:research-design}
One author conducted ethnographically informed research in Beirut, the capital of Lebanon, over the course of two weeks in July 2022. This meant that she (see Researcher Positionalities in Section~\ref{sec:data-analysis}) spent time engaging with people in different settings, including in coffee shops and at public events such as at poetry readings and walking tours of the city. The research was designed to meet people in their everyday contexts; what Brewer~\cite{brewer2000ethnography} calls people's \emph{naturally occurring settings}. We refer to our work as \emph{ethnographically informed} to acknowledge the short-term nature of the fieldwork, and we use \emph{situated security} as a methodological construct to identify and analyse the lived security experiences of Lebanese people. Situated security thus speaks to the ground-up nature of the research design where the ethnographically informed approach situates people's security in their everyday settings and activities. 

Researcher observations and interactions with people in Beirut as well as reflections on such encounters were recorded as field notes and brought into the analysis as discussed in Section~\ref{sec:data-analysis}. These notes took the form of conscious and detailed recordings of the research sites and interactions, which are referred to as \emph{thick description} in ethnography~\cite{geertz2008thick}. 

The researcher stayed in rental accommodation in an area of Beirut, Mar Mikhael, which still shows evidence of the 2020 Port Blast through the many destroyed buildings and dilapidated storefronts. Yet, the area is home to a vibrant community with many LGBTQI+ identifying people having made this area their home. With the researcher being situated in this community specifically and Beirut more broadly, everyday security challenges and technological barriers were felt and observed on a daily basis. The country-wide political and economic collapse had led to significant infrastructural breakdowns which, for example, meant that Internet access was limited as electricity was only guaranteed for two hours a day. The dilapidated sidewalks made it challenging to get to interview locations and it was thus necessary to rely on taxi services. Due to the economic crisis and extortionate currency exchange rates the researcher had to avail of local practices of exchanging currency. This, for example, involved contacting an unknown individual through WhatsApp and exchanging money from that individual's moped.

\begin{table}
  \caption{Overview of interviewed participants in Beirut, the locations and lengths of the interviews.}
  \label{tab:participants}
  \begin{tabular}{cclc}
    \toprule
    ID & Organisation type & Location & Minutes \\
    \midrule
    P0 & Human rights foundation& office& 74  \\
    P1&Human rights foundation& office& 74  \\
    P2& Social change initiative& coffee shop& 39\\
    P3& Political party worker& office& 53 \\
    P4& Digital rights researcher& online & 34 \\
    P5& Internet governance & coffee shop& 41 \\
    P6& Democracy \& digital& office& 71\\
    P7& Digital rights & coffee shop& 46 \\
    P8& Democracy \& elections& office& 93 \\
    P9& Independent media & online& 31 \\
    P10& Diaspora \& human rights& coffee shop& 69 \\
    P11& Digital rights & online& 24\\
    P12& Digitalisation& office& 47 \\
    \bottomrule
  \end{tabular}
  \begin{tablenotes}
  \small 
  \item \small \emph{Notes:} We do not refer to participant IDs in the paper to avoid inferring identification. For a similar reason, we only provide high-level descriptions of the organisation types. P0 and P1 were from the same organisation and took part in the same interview.
  \end{tablenotes}
\end{table}

\paragraph{Participant Recruitment.}
The researcher recruited 13 participants who were knowledgeable about the digital security and socio-political landscape in Lebanon, and conducted 12 semi-structured interviews (nine in person and three online) (see Table~\ref{tab:participants}). Recruitment was done by first identifying potential organisations and individuals who had a public voice on matters relating to digital privacy and security. No further criteria were applied. We emailed identified individuals to explain the research and ask whether they would be interested in participating and/or learning more. Three participants were recruited before arrival in Beirut, while the remaining 10 participants were recruited during the fieldwork either via telephone and/or by visiting their offices. It is perhaps not surprising that the velocity of participation increased significantly during the fieldwork in Beirut, given that the communities and networks of activists and digital rights organisations participating in the research were concentrated in Lebanon. This meant that introductions could be made quickly, with participants introducing new participants and acting as gatekeepers and, to some extent, trust facilitators. The sample size was thus not predetermined, but grew organically as a result of the ethnographically informed approach. 

\paragraph{Interview Process.}
The interviews were semi-structured and followed an interview guide,
which was developed in consultation with Lebanese researchers and recent research outputs by civil society organisations in Lebanon to ensure its relevancy and sensitivity. Existing scholarship was reviewed to understand the gaps within the digital security literature and to inform the theoretical underpinnings. The interviews explored topics related to the wider security contexts in Lebanon including individual perceptions of and experiences with digital security and privacy in the country. The interviews were conducted in English, one of the three main languages spoken in Lebanon, and were audio-recorded with the explicit permission of each participant. Following each interview, the recording was transcribed and anonymised. Once the transcription was complete, the recordings were destroyed prior to leaving the field setting to avoid crossing international borders with potentially sensitive data. 

\paragraph{Ethical Considerations.}
The research received full ethical approval from our institution's Research Ethics Committee and procedures for obtaining informed consent were followed. Additionally, given the political sensitivity of the context within which we carried out the research, one of the key harms that we aimed to mitigate was the potential for participants to become identifiable. Therefore, consent documents emphasised the voluntary nature of the research and the anonymisation procedures we undertook. Data minimisation was prioritised during data collection, transcription, analysis and presentation of the data. This also meant that we did not collect participant identifiers nor do we refer to individuals in the reporting of our findings. This is to further mitigate the potential risk of de-anonymisation at a later date. We also worked with our institutional health and safety office to put in place appropriate risk mitigation protocols. Engaging with the researcher posed minimal risks to participants as Beirut is a diverse city where talking to foreigners is common. The country has witnessed a growing community of Western humanitarian workers within Beirut since the beginning of the Syrian war in 2011 and the displacement of Syrian refugees to Lebanon. Lebanon also has a long history of being a space for collaborative research across the Global North and South.

\subsection{Data Analysis}\label{sec:data-analysis}
While we particularly draw on interview data in this paper, it was the researcher's presence in the Lebanese context that facilitated analysis and interpretation of the situated security of the populations we studied. Thus, our analysis was rooted in our ethnographically informed research design by drawing on ethnography as a methodology that not only shapes data collection but also data analysis by situating it within the context of the data. Transcriptions of the interviews and field notes were compiled into a single data corpus that was thematically analysed using Braun and Clarke's reflexive thematic approach~\cite{braun2019reflecting}. The author who conducted the fieldwork carried out the first cycle of inductive coding of the full data corpus. This was done manually and involved annotating the data, grouping and sorting it into higher-level categories. Through this process, rich descriptions highlighting the researcher's interpretation of the data were recorded. In the second cycle of analysis, the higher-level categories as well as associated descriptions and annotations were presented to the wider research team during a series of collaborative data analysis sessions. During these sessions, the categories were reflexively interrogated, challenged, nuanced, refined and merged (and, indeed, unmerged) in line with individual and collective interpretations among research team members. Through this analytical process the categories were brought into conversation with and interpreted through wider literature on Lebanon and digital security, respectively. Through this process the conceptualisations of ontological security, identity work and positive security also emerged as analytical lenses, enabling a deeper understanding of the intertwining of technological and societal security in the Lebanese context.

\paragraph{Researcher Positionalities.}\label{Res-Pos} 
Our individual positionalities shaped how we interpreted the data. The researcher who conducted the fieldwork identifies as a white female researcher. Despite not having had any prior links to Lebanon, being from Northern Ireland she identified with some of the frustrations expressed by participants when they reflected on their experiences of living in a post-conflict society with fragmented national identities and divides. The research team also included a design researcher who has intimate experience with the conflict, socio-political and economic conditions in Lebanon from the position of a Lebanese woman. She has participated in and researched Lebanese activist spaces from socio-technical and marginalisation perspectives. Finally, the research team also included one researcher who identifies as a woman and uses ethnography to research digital security practices as they relate to at-risk populations. As such the data was analysed from the individual and collective epistemological positions and identities held by the research team~\cite{braun2022starting}.

\section{Findings}\label{sec:findings}
In this section, we first detail the digital security landscape in which our research was situated. We do so in Section~\ref{sec:findings-landscape}. In Section~\ref{sec:findings-digital-state}, we foreground the digital landscape that mediated the relationship between the Lebanese people and the government. In Section~\ref{sec:findings-scapegoating}, we show how those on the margins of Lebanese society -- refugees, LGBTQI+ people and women -- experienced forms of othering. Our findings, while not exclusive to the Lebanese context when viewed individually, collectively demonstrate how gendered and identity-specific notions of exclusion, insecurity and risk are reinforced in Lebanon. In Section~\ref{sec:findings-accountability} we conclude this section by highlighting how the lack of accountability in Lebanon is shaped by a technology sector that has limited situated understanding of the Lebanese context and a corrupt judicial system.

\subsection{Situating Digital Security in Lebanon}\label{sec:findings-landscape}
The digital landscape in Lebanon is situated in the wider post-conflict context, which is shaped by the financial collapse and legacies of corruption and sectarianism. In 2019, the Lebanese economy crashed due to government mismanagement of public finances, a lack of fiscal accountability and sectarian corruption and clientalism~\cite{worldbank2022}. This led to hyperinflation, with the majority of the Lebanese population being forced to live in poverty~\cite{HRW2021b}. During the fieldwork the researcher observed the militarisation of banks and ATMs in Beirut, and read reports of protests which turned violent as personal savings were held from depositors. 

The cost of data had increased exponentially over the preceding year reducing the affordability of accessing the Internet for many people. While data had become more expensive it had simultaneously decreased in availability. Participants reported that the 2-3G coverage across the country, which enabled those who could afford mobile data bundles to access the Internet, had been reduced. This was estimated to eradicate network access for nearly 300,000 people as it limited remote rural connectivity and services for those with lower incomes, including large refugee communities. The variance in Internet provision was compounded by sectarian governance. For example, ministries (e.g., the Ministry of Public Works and Transport) were rotated among the different sectarian parties, granting the benefits of certain services to communities who supported their political agenda. Participants explained that some neighbourhoods also had better access to electricity than others due to the political party who controlled the area. This fluctuation in service provision further led to communities in different geographic areas becoming marginalised.

Households typically had access to WiFi, with some sharing WiFi-related costs with their neighbours or visiting neighbours or public spaces to connect to the Internet. The research revealed how the pooling of resources was a common practice to maintain Internet access. Other research in Lebanon has shown that during times of financial strain, WiFi connectivity would be forfeited as households prioritised paying essential costs~\cite{talhouk2020}. During the fieldwork it was observed that WiFi was available in some public places such as restaurants and caf\'{e}s. However, while research on marginalised groups in Western contexts has shown the importance of publicly accessible digital services and the use of public computers in, for example, public libraries or through case managers or language teachers (see e.g.~\cite{SP:SLIRK18,CHI:ColJenTal18}), public spaces and services did not exist in Lebanon. The research thus revealed how marginalised populations in Lebanon relied on each other for Internet access, rather than government-provided services. Further, public spaces such as public parks where people would traditionally gather in Beirut had been closed down during the 2019-2020 protests and had remained closed. While COVID-19 was said to be the official reason for the continued closure of public spaces, some participants highlighted how such closures were the government's attempt to suppress public upheaval by preventing people from gathering.

Internet access was also restricted by electricity shortages. Electricity bills had continued to increase with inflation rising and reaching 170\% in 2022~\cite{press:Lebanon:FT}, leaving people to prioritise their essential needs. The researcher also experienced this while in Lebanon, where she would ensure to charge all devices when and where electricity was available as only two hours of electricity per day were guaranteed by the government. The researcher spent many evenings in Beirut with flickering lights and no Internet access due to electricity blackouts and noted the normalcy of doing grocery shopping in darkness, using a phone flashlight when the electricity cut off. 

Mobile phones were the primary device used for Internet access in Lebanon, with several prior works also speaking to the role of such devices in refugee communities, e.g,~\cite{salamoun2022,goransson2020}. Yet the complexities of the Lebanese context and the state of digitalisation in the country mean that the level of use and adoption is unclear. Some reports~\cite{Kemp2023} suggest that around 85\% of the Lebanese population are Internet users, yet, what such usage entails and how it is secured remains uncertain. The fieldwork revealed the reliance on WhatsApp as a central form of communication in daily interactions. This was also experienced by the researcher. Despite using service-specific applications such as Bolt for transport and Toters for food delivery, the communication often defaulted to WhatsApp despite the inbuilt chat function on those applications.

The economic crisis has also hindered wider technological advancements in Lebanon, with the expense of new technology soaring. This is exemplified by multiple threads on Reddit~\cite{RedditLaptops} discussing the increased costs of technology in Lebanon such as laptops, due to the devaluation of the Lebanese Lira. In the fieldwork, this was also highlighted as one of the reasons why people had to rely on mobile phones as these could still be purchased at a price that was seen to be affordable to many. The second reason for the slow technological advancement in Lebanon is the emigration of those who have received technological training within Lebanese universities to the Gulf countries or further afield for better opportunities in the face of the economic decline. Articulating this, one participant explained how this was not a brain drain but instead a ``haemorrhage'' that impacted both wider technological skills and security knowledge in the country. 

\subsection{Technological State Control}\label{sec:findings-digital-state}
The digital landscape as it relates to government services is complex and shaped by a government historicised by corruption. One participant explained that the government was purposefully slow to digitalise services because ``the more paperwork the better the corruption.'' However, another participant highlighted that increased digitalisation would not necessarily mean greater transparency within the Lebanese government: ``It is challenging because just like a digital system can ensure governance and efficiency and effectiveness it can be used to hide the cheating also.'' The challenges ingrained in corrupt Lebanese government institutions and services were exemplified by participants with reference to the identity management system in the country. Here, participants noted that while digitalisation efforts had been realised, such as an online appointment booking system for passport renewal and the deployment of biometric passports, the service was overbooked due to legacies of nepotism and lack of staff resources. Further, applying for or renewing a passport still relied on paper-based systems, and a records system that dated back to the initial formation of the Lebanese government. One participant recounted the complexity of registering for government digital systems due to the incoherent approach to registering identity with the State: ``Some people have ID and some do not [\dots] some people do not even have a passport. Even your birth certificate is handwritten.'' Participants collectively highlighted how the corruption and sectarianism that underpinned the post-conflict political system were ingrained in digital government systems and processes.

\subsubsection{Practices of Surveillance}\label{sec:findings-surveillance}
During and since the end of the Civil War (1975-1990), varying levels of control exist in Lebanon~\cite{Salibi1990} with some regions under the informal jurisdiction of different sectarian political parties and experiencing heightened surveillance. Participants explained how they enjoyed certain freedoms in Beirut which were not experienced elsewhere in the country: ``We have to be really honest that certain forms of freedom and certain forms of expression are only possible, for example, in the capital that are not really possible in different parts of the country.'' It was further noted that individuals from regions with tighter socio-sectarian control knew that they were being monitored online and that those of certain sects were more closely monitored. Participants explained that in reality ``people who come from these areas [participant clarified: South of Lebanon, North and Bekaa regions], they are aware that their online presence is being closely policed.'' Digitally enabled practices of surveillance thus extended beyond geographic boundaries that would traditionally have been bounded by sectarian strongholds. However, it is important to note that the Lebanese government does not have technological control similar to that reported in Iran~\cite{grinko2022nationalizing} and one participant highlighted this aspect: ``[they] just do not have the technology for it [\dots] if there were cameras everywhere I would be freaking out right now.'' Further, the government has not, as of yet, exerted control of access to messaging applications and/or online platforms (e.g., Facebook or WhatsApp) that have previously been used to organise movements~\cite{USENIX:ABJM21,ermoshina2022concealing}. However, the sense of being monitored was evident throughout the data with one participant noting: ``I mean people do not trust, there is an assumption that the State has access to all our WhatsApp messages etc. as there is the feeling that the State has access to everything.'' One participant linked the fear of online surveillance to their sense of being insecure, stating that ``everyone is watching everyone and it is not a safe environment'' when referring to digital platforms.

\subsection{Practices of `Othering'}\label{sec:findings-scapegoating}
Marginalisation based on gender has been shown to be intimately linked to the Lebanese sectarian governance systems where State and religious institutions variably regulate sexuality to produce and maintain sectarian political differences (i.e. sextarianism)~\cite{mikdashi2018,Mikdashi2022} and in doing so contribute to acts of othering. This was echoed by participants who emphasised that gender rights were subject to scrutiny by the State and by the multiple religious courts in the country. One participant said sarcastically: ``it is not common to have LGBT rights in Lebanon or women's rights as it is `against' our religions.'' They pointed to how post-conflict sectarianism -- sextarianism -- constrained any potential for meaningful advancement in gender rights and, in turn, contributed to the othering of LGBTQI+ identifying people and women. 

Our data shows that practices of othering occurred on a scale from attributing moral blame, to experiencing abuse and hatred, to being scapegoated. Our findings further highlight how such practices manifested on online platforms such as social media. One participant also explained how these forms of othering were linked to the reduction of socio-economic stability: ``The more the situation deteriorates, the more the marginalised groups get more marginalised [\dots] they [parliament] are not even taking into consideration LGBT rights.'' This correlation between increased othering and economic instability in Lebanon was picked up by one participant who stated that ``in times of crisis things become more protracted and more complicated''. They further highlighted that the majority of people ``are so busy with surviving and securing their economic survival that, for example, a conversation on democracy, on civil rights and on women's rights and domestic [migrant] workers it becomes less important.'' 

One participant suggested that the lack of space for dialogue over civil rights during the economic crisis and due to sectarianism further served the political elite: ``In every society at the time of the crisis the masses need a scapegoat [\dots] there are always attempts by authorities alongside their friends in religious communities to actually stir trouble related to religious issues or moral issues.'' Participants were concerned about the amplified risks that certain groups were experiencing online due to the atmosphere becoming increasingly violent, citing recent `movements' on online platforms against distinct at-risk groups. 

\subsubsection{LGBTQI+ as an `Other'}\label{sec:findings-lgbtqi}
Our findings show how the socio-political and economic situation in Lebanon placed those who identify as LGBTQI+ at a higher security risk. In particular, participants cited how identity was influenced by the patriarchal, sectarian and class-based structure that permeated Lebanese society, while articulating how this resulted in heightened security risks for gender diverse groups. Here, we specifically focus on examples of how such groups experienced online exclusion and threats. The othering of LGBTQI+ identifying people was exemplified by one participant: ``We saw it a couple of weeks ago with the Ministry of the Interior saying `we don't want any activity which is promoting LGBT society', so they are trying always to find a scapegoat.'' Participants further noted how demonstrations calling for LGBTQI+ rights in Beirut, in front of the Ministry of Interior, were met with threats on different platforms: ``Dozens of groups were created online and most of them I would say are [government] intelligence, part of the system, threatening those who are willing to have a sit-in in front of the Ministry of Interior. Threatening them with attacks, beatings and even killing.'' 

Authority-driven incitements of violence became a recurring theme across interviews with participants in Beirut. For example, one participant explained: ``the Ministry of Interior issued a statement accusing the LGBTQ community and not taking a single measure against those who are attacking or threatening [them]'' (we explore this further in Section~\ref{sec:findings-accountability}). The participant spoke for some time on how blaming LGBTQI+ identifying people and the subsequent threats made against them was a means of diverting attention away from the failures of the government: ``These huge offline and online campaigns that were recently launched against LGBT+ community and individuals is very much a political decision, in the sense that it is not really about LGBT but in order to divert the public debate from key issues related to economic livelihood, political rights and political development.'' Societal divisions normalised othering along sectarian lines and established an online environment in which hatred and oppression of LGBTQI+ identifying people was encouraged.

The incitement of violence against LGBTQI+ groups was present in everyday discourse during the fieldwork as well, underpinning the discourses that manifested online. For example during a poetry night one poet (spoken in English) emotionally spoke of their experiences, recounting how when leaving Beirut to visit their family in the South of Lebanon, by dressing differently and leaving their partner behind. They spoke about how these experiences were heightened ``especially now''. While doing a walking tour of the city the researcher documented graffiti relating to LGBTQI+ rights such as one that stated: `the closet doesn't fit us any more.' Yet, when conversing with others regarding such forms of activism, it was observed that since the 2020 port blast -- the explosion of ammonium nitrate in the Port of Beirut that destroyed around half the city -- and with the ever deteriorating economic situation, the glimmer of support for LGBTQI+ people was decreasing. This was recognised in line with the upsurge in scapegoating narratives to divert blame from failing fiscal policies of the State. Insights shared through these observations were furthered by interview participants who reasoned that scapegoating occurred ``because the LGBT community is threatening the system which is very patriarchal and very linked to religious leaders and institutions in Lebanon.'' As emphasised here, participants articulated how LGBTQI+ identifying groups were being viewed as a direct challenge to traditional authority and established moral norms in Lebanon: ``Whenever you threaten the religious and sectarian institutions, they [the institutions] say that the human rights rhetoric and LGBT rhetoric are imported by Westerners.'' This was a common point of reflection for most interactions in Lebanon when talking about gender identity and marginalisation.

Participants also shared how othering was intimately tied to sectarian geographic strongholds in Lebanon and Beirut more specifically, with some neighbourhoods deemed safer than others for LGBTQI+ people. One participant stated: ``You would see them [the political wings of Lebanese Christian groups] having big beards wearing a cross, and whenever they see a LGBT person even walking in the streets, they will attack him in the streets just because he is walking in their area.'' These experiences of heightened LGBTQI+ targeted violence in certain neighbourhoods also manifested on social media. Platforms such as Facebook were described as further enforcing the ``polarisation'' of Lebanese society and was a space said to be rife with hatred targeted towards LGBTQI+ identifying individuals. One participant spoke about the challenges faced when liking social media posts related to LGBTQI+ groups and the backlash which often resulted in threats to personal safety: ``They [LGBTQI+ groups] became silent as it has become violent, very very violent on social media, but also on the ground people wanted to protest and other religious groups saying they were preparing the guns.'' One participant recounted the experiences of a well-known drag queen: ``We have a drag scene in Lebanon and we have a famous drag queen [\dots] she will get lots of comments, hateful speech, she will be constantly reported and her account will be taken down and she experiences censorship by the platform, by people who are working to censor her.'' Occurrences of LGBTQI+ related othering were particularly amplified online: ``I could be sitting here and my information ecosystem could be telling me `oh you know there are more gays in Lebanon and you know they are getting money from NGOs to be gay and they are corrupting us'. So if the algorithm keeps promoting that, what happens?''. The speculative concern of this participant highlighted the potential danger of algorithms in promoting scapegoating and blaming through inciting fear of those who posed a challenge to traditional moral norms during a time of socio-economic decline.

\subsubsection{Amplified Scapegoating of Refugees} \label{sec:findings-refugees}
The link between othering, specifically scapegoating, and amplified risks experienced by those on the margins of Lebanese society was further seen in the case of refugees. Lebanon is home to several refugee groups including Palestinian, Iraqi and Syrian refugees who have also been impacted by sectarianism~\cite{masalha2002secretarianism}. For example, the naturalisation of Palestinian refugees that fled the Israeli-Palestinian conflict was and still is a contested topic in Lebanon with Maronite Christians fearing that granting citizenship would skew the demographics and, in turn, the power dynamics towards Muslims~\cite{knudsen2009widening}. Tensions surrounding naturalisation resulted in Lebanon not signing the 1951 Convention related to refugees~\cite{knudsen2009widening,masalha2002secretarianism}, leaving the majority of refugees living in poverty due to the systematic social and legal exclusion and discrimination restricting their access to social and occupational institutions~\cite{hanafi2012social,janmyr2016precarity}. Research has shown how such policies contribute to the othering of refugees within Lebanese society, limiting their access to health services~\cite{talhouk2016syrian} and food aid~\cite{talhouk2022refugee,talhouk2020food}.

While our research did not directly engage with refugees, participants voiced their concerns about the current politico-economic situation and how they witnessed refugees being placed as scapegoats for the many government failures: ``I am being pessimistic about it, I feel like there might be some attacks, managed attacks, on the Syrians for example.'' Participants noted the correlation between the deteriorating political and economic situation and the amplified security risks for refugees: ``The son in law of the President [and leader of a political party] [\dots] has been saying the Syrians are the problem, and the reason for the economic crisis because they took your money outside of the country.'' This form of politically driven scapegoating and assigning of blame to a particular refugee group was amplified through social media platforms where video snippets and quotes from this occurrence were widely shared. Participants emphasised how the digital security risks for these groups were amplified through their presence on online platforms. However, accessing digital content also enabled refugees to maintain social connections. This was underpinned by one participants who spoke about the wide use of mobile phones and social media among refugee groups: ``Almost everyone, even someone who lives in a [refugee] camp, will have access to a smartphone and will have Facebook, Twitter etc. downloaded.'' Existing security-focused research with refugees has further shown how the mobile phone plays an important, yet doubled-edged, role for refugees, being both a security enabler and a security risk~\cite{CHI:ColJenTal18}.
 
Participants further spoke of how politicians diverted blame for the economic crisis to refugees through powerful and fear-based discourses: ``They need an outsider to scare people off and our politicians are very good at scaring people off.'' This was felt to such an extent that one participant explained how this was an online ``campaign'' against Syrian refugees, highlighting the organised nature of this rhetoric, especially around food insecurity. Further examples were found on social media where senior Lebanese politicians were observed to fuel the scapegoating of refugees. They did so through narratives of blaming refugees for the financial collapse by suggesting, for example, that their own salaries were less than what a Syrian refugee would receive from aid organisations. Participants also pointed to another online campaign which focused on subsidised wheat from the UN being given to refugee communities rather than Lebanese populations. This campaign was explained to be used to divert attention away from government corruption and the smuggling of wheat by the political elite in Lebanon. As noted by one participant: ``What many civil society organisations have found is that actually algorithms tend to amplify content that is inflammatory because that is what gets reactions.'' Our findings also exemplify how social media posts, as part of politically driven scapegoating campaigns, went viral given the political power driving them. This amplified the digital security needs of marginalised groups.

While there was a general consensus that blame should not be placed on refugee groups, they explained how this form of scapegoating spoke to the pressures felt by many in Lebanon that were amplified by the scarcity of economic resources. One participant highlighted this when they asked: ``If I am not able to sustain you [referring to the Lebanese people] how am I meant to sustain some stranger [refugees]?''. Class was also seen to override non-citizenship status: ``The very rich Syrians who sit with really rich Lebanese will say `oh you know those poor refugees, they don't see themselves as this [refugees] because they have means, they have access, they have possibilities to leave etc.' [\dots] You have the differences I think more in the middle [class].'' This highlights how economic tensions were a dividing factor, above ethnicity, furthering the scapegoating of `poor' refugee communities. The division of `rich' and `poor', `us' and `them' re-emphasises how refugees were seen as `the other'. 

\subsubsection{Silencing of Women}\label{sec:findings-women}
The `sextarian' divisions~\cite{mikdashi2018,Mikdashi2022} were said to be enabled because Lebanese society is ``very patriarchical''. This was exemplified by one participant who highlighted that as socio-economic conditions declined, certain groups would face greater online bullying and censorship: ``LGBT people are going to face problems and Lebanese women will be scorned a lot.'' Another participant explained that women were the target of attacks, because ``the level of sexism is extremely high along with sexual harassment both online and offline.'' One participant highlighted how women who were perceived to be of a certain socio-economic class would experience amplified online security risks: ``As well as being a patriarchal society it is a very classist society.'' The participant relayed the case of a Lebanese singer who received a lot of online hatred due to the way she dressed and how she applied her makeup which was seen to signify belonging to a particular class, they concluded: ``all of these things play out online.'' The threats faced by Lebanese women in public life were further articulated by one participant: ``There is a journalist, a Shia woman who is very critical [of the government]. Everything she says is relentlessly criticised until it becomes a trend on Twitter saying she is a liar.'' 

Another example was given of a woman doctor who was successfully providing sex education and female medical advice on TikTok. Following a campaign by religious institutions calling on people to report her, she was banned on social media platforms. One participant noted that ``she kept getting banned over and over and over again'' due to the organised silencing campaign. In some cases, online reporting and censorship of women led to them being taken in for government investigation. For example, a well-known female comedian protested online against COVID-19 restrictions that required providing internal security forces with details about when and why she wanted to leave the house, sharing online that she needed to buy period products. One participant explained how she was ``joking to the soldiers about whether they're going to buy her pads or her tampons''. This led to her being called for investigations at a military court under the charge of insulting the security forces.

Participants also commented how higher-risk groups in the Lebanese context, particularly women, often engaged in ``a lot of self-censorship'' as those who spoke out about human rights issues experienced ``shadowbanning.''\footnote{\emph{Shadowbanning} refers to being blocked from social media without knowing that one has been blocked or that one's comments are not visible.} This was noted about women who publicly took a stand against the sectarian leaders (participants mentioned stand-up comedians, for example) and would experience that their content ``is not being shown as much''. One participant explained how they ``will call out being shadowbanned. Or they will post a photo of themselves and then put important information in the text as a photo of a happy couple or whatever will trick the algorithm.'' Thus, participants pointed to distinct and sometimes subtle mitigation strategies to circumvent being shadowbanned. 

Despite the digital security threats experienced by those marginalised in Lebanon, participants explained how taking online risks was a necessary form of education. One participant shared the pride they felt when they saw younger people sharing something ``super feminist about women being under attack in the region'' and feeling ``grateful to social media'' as many young people in Lebanon were unable to access progressive politics within their families. Online platforms were also seen as a ``tool for accountability'' for some, with participants citing intersecting issues of class, race and gender being raised online. As one participant noted: ``Maybe it will be around a beach resort who ban migrant workers from swimming, and we will come out [online] and say you [the resort] are being racist we are boycotting this place and then they will put a clarifying statement.'' This highlights the dual role that social media platforms played in the Lebanese context, e.g, as spaces for advocacy and for amplified marginalisation.

\subsection{Situating Accountability}\label{sec:findings-accountability}
Our findings illustrate the digital landscape in Lebanon and how this is enmeshed in societal conditions shaped by the post-conflict Lebanese context. While the participants in our study articulated how the Lebanese authorities exerted technological control and monitoring (particularly in some sectarian strongholds), they also gave several examples of how digital systems were either broken or not designed for the post-conflict challenges and cultural norms of Lebanese society. Further, and most prominently, our findings brought to the fore the interwoven practices of othering experienced by marginalised populations in Lebanon; forms of scapegoating and silencing that, whilst situated in the politico-economic crisis of post-conflict Lebanon, often manifested online. In Section~\ref{sec:discussion} we discuss possible digital security responses, while foregrounding how designing with positive security may give rise to hopeful technologies that aim to foster dialogue and reconciliation. We conclude our findings by exemplifying how the lack of accountability in both the private and public sector allows for extensive forms of othering to continue within the Lebanese context, and how this is underpinned by sectarianism and a lack of situated security understanding.

Participants highlighted how current efforts by the private technology sector to mitigate online forms of violence were tokenistic and a ``public relations face''. They articulated how such efforts lacked cultural sensitivity, making them ineffective. One participant explained how social media platforms, in this case Facebook, did not have a system which could sensitively address digital risks experienced by Lebanese populations: ``They'll hire someone who looks like you, a brown woman with curly hair and she will be Palestinian and my colleagues or whoever will say this [online harm] is outrageous and she will reply in a Palestinian accent and say we are looking into it [\dots] then she goes back inside the company and she isn't able to do anything.'' The lack of situated cultural understanding was also noted in how the positioning of employees to represent a region within global technology offices would not provide sufficient security to local communities. This was particularly the case in Lebanon where participants explained that Lebanese dialects were different to dialects in the region and the Standard Arabic Dialect. When compounded by ongoing sectarian divisions, distinguishing between what was inflammatory and/or harmful content without local cultural knowledge was explained to be impossible. With respect to shadowbanning (Section~\ref{sec:findings-women}), participants also highlighted that ``there is no transparency, which is the main problem when it comes to corporate accountability because when we go to them and say `you are shadowbanning stand-up comedians in Lebanon' [\dots] they will ask, how, why, how do you know?''.

Further highlighting the significance of situating technology design in the contexts they are being implemented and used, participants gave examples of how simply adopting standard Western-based digitalisation processes often led to tensions between cultural aspects and customs, and the implementation of digital systems. For example, digital systems developed to cater to Western norms were explained to be incompatible with the multiple paper-based identity systems in Lebanon. One participant highlighted with reference to Arabic names: ``Your name, how it is written, how it could be written in so many variations [when translated from Arabic to other languages]; so in one [government] system it could be different to another [\dots] because of the lack of [consistent] identity [documents] you might find different variations of my name.'' Such inconsistencies required identity verification processes with which Western-based systems struggled, as the verification often needed to extend beyond the individual to include data such as parents' names and place of birth or even uncles', grandfathers' and great grandfathers' identity details. These insights speak to the findings of~\cite{SP:SLIRK18}, where the authors highlight how authentication mechanisms, such as password recovery, are rooted in Western cultural norms.

Participants highlighted that within failing governance structures and tokenistic private sector systems, there was no expectation that those who engaged in online abuse would be held to account by the Lebanese authorities: ``The authorities never protect victims [of] violence, they are always here to protect the perpetrators of violations and to perpetuate a culture of no accountability whatever happens.'' This lack of accountability extended to digital security where the jurisdiction and processes of government agencies such as the Cyber Crime Bureau (CCB) were explained to be opaque at best, with one participant noting how the CCB could ``summon you for what you have posted online''. One participant further explained the role of the CCB: ``This is an attempt at or an attempt to curtail freedom of speech but if you have enough clout and you have the money to hire a lawyer or belong to a certain segment of society it doesn't affect you in the end.'' More broadly, our findings suggest that the corrupt Lebanese judicial and political system also meant that there was a lack of digital security and privacy policies in place to protect human and digital rights. Thus, the the lack of prioritisation and accountability of/for digital security was directly tied to the general lack of accountability in the country. This is supported by Salloukh~\cite{Salloukh2019}, who shows how governance in Lebanon can be characterised as a fragmented system of clientalism and corruption with no accountability. 

\section{Discussion}\label{sec:discussion}
Here, we outline our study's key takeaways. In Section~\ref{sec:dis-moral-panics} we focus on digital security of and for marginalised groups in the Lebanese post-conflict context, showing parallels with broader digital security research focusing on higher-risk populations. In Section~\ref{sec:dis-post-conflict} we draw on notions of positive security to support our argument for the need to design security technologies that enhance dialogue across fragmented and post-conflict societies. We conclude our discussion in Section~\ref{sec:dis-recommendations}, where we set out implications for digital security research.

\subsection{Importance of Digital Security in Lebanon}\label{sec:dis-moral-panics}
Marginalised groups being othered during times of political and economic instability is not a new phenomenon as we highlight in Section~\ref{sec:rw-scapegoating}. Yet, our findings show that marginalised groups in Lebanon are positioned as the drivers of societal-level failings as they are seen to challenge the traditional morals that protect the ontological security of the regime. This transfers blame from the political leadership to groups who are already marginalised, heightening their at-risk position in society. By framing such groups as an `other', the Lebanese political leadership also constructs and reinforces its own identity as the (legitimate) ruler of Lebanon. Our findings show how such practices manifest in online contexts in Lebanon, leading to heightened and amplified digital security risks for already at-risk populations. While this is not in itself a matter that can -- or indeed should -- be solved through technological interventions, we argue for a digital security that is situated in the context of people and their daily lives. As our findings have shown, practices of othering as experienced by refugees, LGBTQI+ identifying people and women in Lebanon (Section~\ref{sec:findings-scapegoating}) speak to the ontological insecurity of these groups as well as the need to protect technology, its availability and security.

Our findings show the need for digital security literature to consider intersecting forms of marginalisation and political, economic, social and cultural structures that create the conditions for high levels of insecurity. Not doing so ignores the drivers of key vulnerabilities and threats and, thus, contributes to the further marginalisation of populations who are especially impacted; those othered in the Lebanese context. We thus echo similar calls made by security researchers working with marginalised groups, e.g.~\cite{USENIX:GHRR22,USENIX:SCBAPB22,SP:SLIRK18}, who have shown how the conditions of marginalisation directly impact the digital security threats experienced by these populations and that such digital security threats are neglected in the literature~\cite{USENIX:SCBAPB22}.

Our study raises the question of what kind of digital security agenda is required to better cater to the needs of marginalised populations whose threat horizon is amplified through digital technology. In the Lebanese context, State technological control and online surveillance of particular groups that are seen to challenge societal norms amplifies the digital security risks of those populations (Section~\ref{sec:findings-digital-state}). Our work shows that marginalised populations in post-conflict Lebanon are not free from persecution by the State nor can they use technology freely. While presenting different challenges to digital security, broader literature on the computer security needs of higher-risk populations also points to more situated security technologies. This includes security-focused work with refugees and migrants~\cite{SP:SLIRK18,CHI:ColJenTal18,CHI:ColJen19,CSCW:TCJBGGAAM20}, LGBTQI+ identifying people~\cite{CHI:GMSMTS18,CHI:LHKZH20,USENIX:GHRR22} and migrant domestic workers~\cite{USENIX:SCBAPB22}, as we highlight in Section~\ref{sec:rw-higher-risk}. What these studies show is how such populations mitigate their digital security risks by relying on trusted relations and routine activities to establish ontological security. In our study, this was for example evident in how those without Internet access or electricity would visit neighbours to use their WiFi and charge their mobile phones; or how refugees would remain active mobile phone and social media users to re-establish routine practices and connections that kept them connected to their homeland (Section~\ref{sec:findings-landscape}).

Participants in our study noted how during the recent economic crisis, digital security risks were increasingly intertwined with offline risks. For example, the online scapegoating of LGBTQI+ people and the silencing of women were often accompanied by physical and State violence. Those responsible were said to not be held accountable by the Lebanese authorities or through accountability infrastructures such as the courts (Section~\ref{sec:findings-accountability}). The intersections of online and offline security risks are foregrounded throughout our findings. This intertwining brings technical, social, political and cultural notions of security into conversation. We now turn to how digital security scholars might respond.

\subsection{Positive Security for Post-Conflict Settings}\label{sec:dis-post-conflict} 
In this section we discuss how designing within a positive security framework, which we outline in Section~\ref{sec:rw-ontological-security}, presents an opportunity for security researchers to work with post-conflict communities to co-create hopeful security technology. Drawing on an understanding of positive security as ``the freedom to live free from fear''~\cite{mcsweeney1999security}, we call for a digital security that not only starts from the goal of countering existing threats, but which focuses on enhancing dialogue across fragmented societies. This speaks to the role that technology has been shown to have in post-conflict transitions (e.g.~\cite{martin2018peacekeeping,limani2018challenges}). In the context of Lebanon, designing for positive security draws on research that emphasises the need to design across the multi-life span of populations that have experienced conflict. This work highlights the importance of accounting for historically rooted, conflict-related fragmentation in a manner that enables transitioning towards peace~\cite{yoo2016multi}. This approach brings into conversation the fragmented identities inherited from the Lebanese Civil War to work towards a reconciliation of society~\cite{yoo2016multi}; disrupting the `us' and `them' rhetoric that we show underlies the mechanisms of othering in Lebanon. Here, digital security researchers can look to technologies designed for post-conflict reconciliation such as digital memorials~\cite{durrant2014human}, and research that calls for a re-orientation towards designing for the desire of those marginalised to achieve meaningful inclusion~\cite{Toyama2017,Toyama2020,CSCW:WGTD21} and a participatory approach that actively designs across societal and sectarian divides~~\cite{talhouk2022dialogues}. This is important as our findings show how the lack of space for dialogue further served the political elite (Section~\ref{sec:findings-scapegoating}).

\paragraph{Digital memorials.} Digital memorials range from digital headstones to collective online spaces to share stories in commemoration. The authors of~\cite{friedman2017multi} note how curators of digital memorials in post-conflict settings act as stewards of stories of conflict, designing memorials underpinned by principles of accuracy, credibility, transparency, safety and security. We ask digital security researchers to collaborate with human rights groups and act as stewards of narratives of othering such as those presented in our work. Transparently curating narratives of othering that unfold online would enable credible accounts of violence to be presented back to current and future generations of Lebanese, in support of reconciling historically rooted violence. Digital security researchers are well placed to undertake this work given their expertise in formulating and navigating digital security safeguards~\cite{leerssen2023end}, such as the shadowbanning mechanisms reported by participants in our work (Section~\ref{sec:findings-women}). This work would make visible the narratives that would otherwise be hidden within Lebanon's post-conflict history. Van Ommering and el Soussi~\cite{van2017space} show in their work on the digital memorial for the estimated 17,000 people who went missing during the Lebanese Civil War how such memorials ``open up spaces that remain closed in the offline world, enabling survivors to share their stories, build collectives, demand recognition, and advocate for justice.'' Yet, key challenges for security remain. Digital memorials hold sensitive data about individuals, their stories, families and identities, while not designed with security in mind. Thus, the development and maintenance of secure digital platforms for such memorials while safeguarding accessibility and ownership are important matters for security in this context.

\subsection{Implications and Future Research}\label{sec:dis-recommendations}
Here, we concretise the implications of our findings for digital security researchers and practitioners, beyond positive security.

\paragraph{Situating digital security in post-conflict settings.} Post-conflict contexts bring to the fore distinct digital security challenges, particularly for already at-risk groups as our findings on LGBTQI+ identifying people, refugees and women show. A wide range of existing security-driven work with higher-risk populations such as, for example, journalists~\cite{USENIX:MCHR15}, activists and protesters~\cite{USENIX:ABJM21,SP:DSKB21} (see also Section~\ref{sec:rw-higher-risk}) has aimed at protecting such populations against (State) surveillance and censorship while mitigating security risks. Speaking to this body of work, we argue that post-conflict societies should also be given particular attention within the security community. We do so because of the many, intertwining and continuous threats to security, safety, privacy and identity as well as economic and political risks that shape people's lives in the `transition continuum'~\cite{brown2011typology}, sometimes for decades and across generations. At a technological level, we suggest working towards specific security choices and controls for people in post-conflict societies, building on existing protection mechanisms against politically motivated attacks and surveillance. Situating digital security in post-conflict settings moves beyond advocating for specific technological interventions, however. Here, computer and social scientists need to work together to uncover and understand the social foundations of the security technologies that are relied upon for protection. 

\paragraph{Security misconceptions.} Our findings indicate some security misconceptions among participants. Participants suggested that the Lebanese State had access to, e.g., WhatsApp messages, indicating a misconception about the promises of end-to-end encryption. The lack of visible surveillance technology present in Lebanon in comparison to neighbouring states also led some participants to question the ability of the Lebanese authorities to monitor interactions. Yet, our findings reveal less overt -- and less reliant on technology -- mechanisms of surveillance that took place between individuals and communities as an extension of the State (Section~\ref{sec:findings-surveillance}). Coupled with limited digital security expertise in Lebanon due to the emigration of those with technology and/or security education (Section~\ref{sec:findings-landscape}), there is an immediate need for digital security information and education in the country. Our work further suggests how culturally-rooted misconceptions held within social media companies has meant that what might be considered harmful content in the Lebanese context is missed (Section~\ref{sec:findings-accountability}). Others have also shown how common security practices such as password creation and related security questions rely on specific cultural (Western) knowledge~\cite{SP:SLIRK18}. To bridge not only technical and social knowledge, we advocate for collaborative security-driven research across the Global North and South, where researchers work together to formulate research problems and responses. As we note in Section~\ref{sec:research-design},  Lebanon has a long history of being a space for collaborative research across the Global North and South. 

\paragraph{Shadowbanning.} Participants commented how those who spoke out against the sectarian leadership experienced shadowbanning (Section~\ref{sec:findings-women}), resulting in a lack of counter narratives to the upsurge in scapegoating and government-affiliated online groups (Section~\ref{sec:findings-lgbtqi}). Our research shows how the concealing of specific content amplified the silencing and scapegoating of marginalised groups in Lebanon. Hence, in the Lebanese context, the practice of shadowbanning across platforms is not a safeguard -- quite the contrary. Our work thus contributes to an emerging body of scholarship pointing to the harmful effects of shadowbanning on marginalised groups~\cite{CSCW:HDNW21,middlebrook2020grey,Nicholas22}. Bloch~\cite{bloch2021content} further highlights how States (through law enforcement) are increasingly influencing the governance of social media content. We caution against developing bespoke moderation policies and tools for the Lebanese context without further situated security research. Yet, we urge social media companies to work with security researchers, across the Global North and South, \emph{and} populations in post-conflict contexts, to situate their technology development and content moderation approaches. Echoing Nicolas~\cite{Nicholas22} we also call on social media companies (Facebook being the most prominent in our findings) to publicly and transparently report their moderation practices, enabling researchers to examine the impact of shadowbanning on at-risk populations.

\paragraph{Mesh messaging.} Our findings demonstrate the unreliability of Internet access for those living in post-conflict Lebanon. This was underpinned by the rising costs of data, a reduction in 2-3G provision leaving rural and poorer communities without Internet access (Section~\ref{sec:findings-landscape}) and growing State surveillance (Section~\ref{sec:findings-surveillance}); coupled with the pooling of resources (e.g., the sharing of WiFi) within communities and high levels of mobile phone use among marginalised populations (Section~\ref{sec:findings-landscape}). These conditions represent key barriers for secure Internet-enabled communications and thus make the Lebanese post-conflict context a prime use-case for mesh network applications that provide communication capabilities over Bluetooth. Indeed, the market leader in this space, Bridgefy~\cite{Bridgefy2023}, is often promoted for use in situations of social unrest and Internet blackouts (e.g,~\cite{Bridgefy2021}). However, recent work has shown devastating security vulnerabilities in the application~\cite{albrecht2021mesh,albrecht2022breaking}. While the level of actual adoption of this technology is also limited~\cite{USENIX:ABJM21}, the spikes in downloads from within areas that witness conflict\footnote{Most recently from within Ukraine~\cite{Bridgefy2022}.} suggest a need for offline messaging that is not provided by existing messaging applications.  While other solutions exist (see~\cite{albrecht2021mesh}) none cater to at-risk people in post-conflict settings who require secure, offline, easily accessible and usable (mass adoption being a key criteria for a mesh network) and reliable solutions. This remains a pressing security question. Future mesh development work might involve grassroot groups to enable, as the authors of~\cite{aouragh2015fcj} note, ``security engineers [\dots] to step into the language of collective action within a political project'' to develop security tools that meet the needs of the populations they aim to serve.  

\paragraph{Ethnography.} We echo the authors of~\cite{USENIX:SCBAPB22}, who call for security researchers to expand the methods used to explore the security practices of populations at the margins. In~\cite{USENIX:ABJM21} the authors call for security research grounded in ethnographic methods to uncover what higher-risk populations take for granted. While we adopted an ethnographically informed approach, our work did not allow for more extensive immersion in these communities due to the two-week fieldwork. Future work should thus consider longer-term ethnographic work in post-conflict settings to deepen the digital security insights that we have begun to uncover here. Our fieldwork was further informed by the involvement of digital and human rights organisations in Lebanon. This was important for our understanding of the political and legal landscape in the country. We suggest that digital security researchers seek out collaborations with interest groups, both as informants and interlocutors (or gatekeepers) to access often hard-to-reach settings and participants. Other security researchers have sought such collaborations. In~\cite{USENIX:SCBAPB22} the authors partnered with Voice of Domestic Workers for their work with migrant domestic workers. In the context of security research with refugees, in~\cite{SP:SLIRK18} the authors engaged case managers, while in~\cite{CHI:ColJenTal18,CHI:ColJen19} the authors worked with language teachers. In all cases, such interlocutors enabled the research to sensitively uncover distinct security needs and practices of such populations.

\section{Conclusion}
Through an ethnographically informed approach our research highlights the interwoven historical, social and political factors in the post-conflict context of Lebanon and how this in turn amplifies digital security risks for marginalised populations during times of political instability. This is caused by the strongly rooted sectarian divisions. We show how populations on the margins -- particularly LGBTQI+ identifying people, refugees and women -- are targeted through practices of othering on a scale ranging from online silencing, to online targeted abuse and hatred, to outright scapegoating. We discuss how the threats facing already at-risk populations are heightened during times of crisis and amplified through digital technology. In doing so, we concluded our discussion by arguing for design interventions rooted in a positive security framework, where technologies are designed to enhance dialogue and reconciliation within post-conflict contexts.

\paragraph{Limitations}\label{sec:limitations}
First, the fieldwork took place over two weeks in Beirut and thus has the potential to be expanded both in time, scope and geography. Second, interviews were carried out with people who had a leading voice in political debates in Lebanon. The work can be expanded to work with different societal groups who are not `experts' in the fields of digital and human rights. Third, not all participants had direct experience of being othered, yet all participants could discuss the practices of othering as being situated in the post-conflict context of Lebanon. Future work on othering should consider engaging those at the margins of Lebanese society. Fourth, the level of participation has limitations such as language barriers. While one of the authors speaks Arabic, the fieldworker does not, however English along with French is widely spoken in Lebanon. Finally, there is an inherent bias in research using interviews and focusing on security and/or politically charged topics, given that participants self-select. It could be that some participants decided against participation as a result. We tried to overcome this limitation by immersing ourselves in the setting through an ethnographically informed approach.

\section*{Acknowledgment}
This work would not exist without the many contributions from people, both in and outside Lebanon, who so generously gave of their time to speak to us and who have remained engaged with the research throughout. We thank the anonymous reviewers for their constructive feedback. The research of McClearn was supported by the EPSRC as part of the Centre for Doctoral Training in Cyber Security for the Everyday at Royal Holloway, University of London (EP/S021817/1).

\bibliographystyle{plain}
\bibliography{local}

\end{document}